\documentclass[3p,times,procedia]{elsarticle}
\usepackage{nupha_ecrc}

\volume{00}

\firstpage{1}

\journalname{Nuclear Physics A}

\runauth{}

\jid{nupha}

\jnltitlelogo{Nuclear Physics A}

\usepackage{amssymb}
\usepackage{amsmath}
\usepackage[figuresright]{rotating}

\newcommand{\trento}{T\raisebox{-0.5ex}{R}ENTo}
\newcommand{\avg}[1]{\langle #1 \rangle}

\newcommand{\T}{\tilde{T}}

\newcommand{\x}{\mathbf x}
\newcommand{\y}{\mathbf y}

\newcommand{\tran}{^\intercal}

\newcommand{\st}{_\star}
\newcommand{\ex}{_\text{exp}}

\begin{document}

\begin{frontmatter}



%% Instructions from Editor: Please use the following \dochead only in the preprint version (e-print arXiv etc.); 
%% use empty \dochead{} when submitting to Nuclear Physics A!
\dochead{XXVIth International Conference on Ultrarelativistic Nucleus-Nucleus Collisions\\ (Quark Matter 2017)}
%% e.g. \dochead{17th International Conference on Dynamical Processes in Excited States of Solids}

\title{Determination of Quark-Gluon-Plasma Parameters from a Global Bayesian Analysis}


\author{Steffen A. Bass, Jonah Bernhard and J. Scott Moreland}

\address{Department of Physics\\ Duke University\\ Durham, NC 27708-0305}

\begin{abstract}
The quality of data taken at RHIC and LHC as well as the success and sophistication of computational models for the description of ultra-relativistic heavy-ion collisions have advanced to a level that allows for the quantitative extraction of the transport properties of the Quark-Gluon-Plasma. However, the complexity of this task as well as the computational effort associated with it can only be overcome by developing novel methodologies: in this paper we outline such an analysis based on Bayesian Statistics and 
systematically compare an event-by-event heavy-ion collision model to data from the Large Hadron Collider.  We simultaneously probe multiple model parameters including fundamental quark-gluon plasma properties such as the temperature-dependence of the specific shear viscosity $\eta/s$, calibrate the model to optimally reproduce experimental data, and extract quantitative constraints for all parameters simultaneously. The method is universal and easily extensible to other data and collision models.
\end{abstract}


\end{frontmatter}

% \linenumbers

\section{Introduction}
\label{}

Relativistic heavy-ion collisions produce a hot, dense phase of strongly-interacting matter commonly known as the quark-gluon plasma (QGP), which rapidly expands and freezes into hadrons \cite{Arsene:2004fa,Adcox:2004mh,Back:2004je,Adams:2005dq,Gyulassy:2004zy,Muller:2006ee,Muller:2012zq}.
Since the QGP is not directly observable -- only final-state hadrons are detected -- present research seeks to quantify the fundamental properties of the QGP, such as its transport coefficients and the nature of the initial state, through comparisons of experimental measurements to computational model calculations.

Computational models must take a set of input parameters including the physical properties of interest, simulate the full time-evolution of heavy-ion collisions, and produce outputs analogous to experimental measurements.
The true values of the physical properties are then extracted by calibrating the input parameters so that the model output optimally reproduces the experimental data.
This generic recipe is called ``model-to-data comparison''. Challenges faced by this type of analysis are the amount of computational effort required to scan the parameter space of the model and correlations among the input parameters which may affect multiple observables, so that they cannot be constrained independently. 

This article introduces the use of a Bayesian model-to-data analysis for the extraction of QGP properties such as the temperature-dependence of its specific shear viscosity. It is designed to serve as an example of many different possible applications. Due to space constraints, only a broad sketch of the analysis can be given -- for a rigorous description we refer the reader to \cite{Bernhard:2015hxa,Bernhard:2016tnd}.

\section{Bayesian Model to Data Comparison}

\begin{figure}[thb]
\begin{center}
\includegraphics[width=0.98\linewidth]{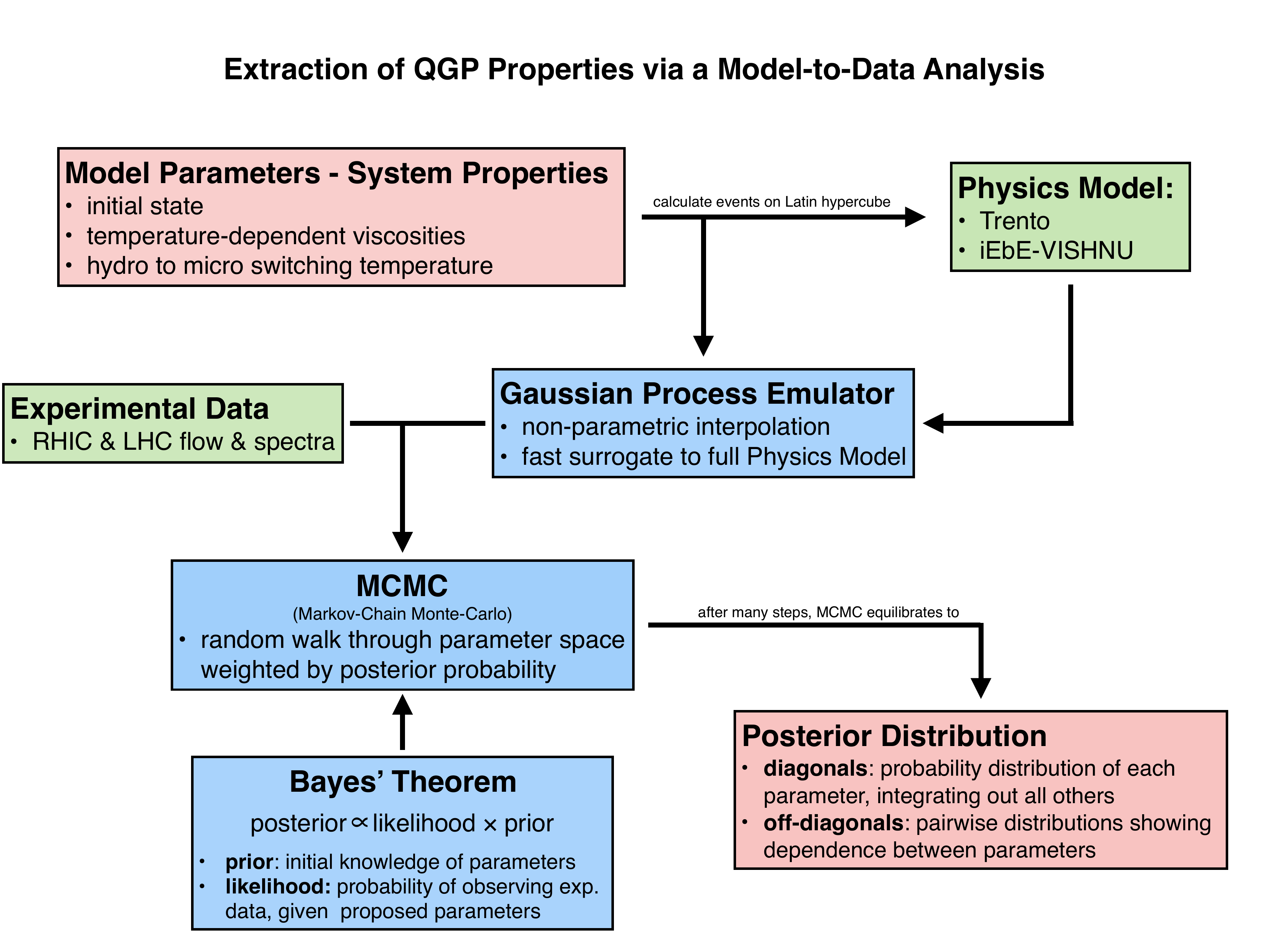}
\caption{\label{flowchart}
Schematic overview of a Bayesian Model-to-Data analysis.
}
\end{center}
\end{figure}

Figure~\ref{flowchart} provides a schematic overview of the of the different components of a Baysian model-to-data comparison. Starting point is a computational physics model with a set of model parameters encoding the physics that one wishes to extract from the data. Here, we use a full event-by-event heavy-ion collision model, VISHNU, based on relativistic viscous fluid dynamics with a microscopic hadronic afterburner \cite{Shen:2014vra,Moreland:2014oya,Bass:1998ca}. The model has 12 input parameters that encode the initial condition, temperature dependent shear- and bulk viscosities and a couple of additional quantities, such as the thermalization time of the QGP and the transition temperature from the hydrodynamic evolution to the microscopic evolution. We calibrate to multiplicity and flow data from the Large Hadron Collider (LHC) -- for the sake of simplicity we focus here on data taken by ALICE at  2.76 and 5.02 TeV beam energy respectively \cite{Abelev:2013vea,ALICE:2011ab}. 

In principle the model can be evaluated on a fine grid of points in its 12-dimensional parameter space. One can then utilize algorithms such as Markov Chain Monte Carlo (MCMC) to rigorously explore the complex high-dimensional parameter space. However, performing the MCMC analysis requires a very large number of model evaluations in parameter space -- often thousands or millions, depending on the problem at hand.
Heavy-ion collision models may run for several hours, so a direct MCMC approach may require in excess of $10^{14}$ CPU hours and is thus intractable.
The situation is exacerbated when studying event-by-event fluctuations as opposed to average quantities:
while event-averaged models save computation time by using a smooth initial condition and single hydrodynamic calculation, event-by-event models have realistic, fluctuated initial conditions, each of which requires its own hydrodynamic treatment.
Many thousands of complete events are necessary \emph{at each point in parameter space} to capture event-by-event fluctuations.

These limitations may be overcome through a modern Bayesian method for analyzing computationally expensive models \cite{OHagan:2006ba,Higdon:2008cmc,Higdon:2014tva}.
A set of salient model parameters is chosen for calibration -- the set should include any fundamental physical properties of interest -- and the model is evaluated at a relatively small number of points that cover the full parameter ranges of interest. The selection of the points at which to evaluate the model is done via a Latin hypercube algorithm that ensures a full and smooth coverage of the entire parameter space.
Once these design points are chosen, the full physics model is run at these points and Gaussian process emulators are trained \cite{Rasmussen:2006gp} on the observables calculated from the model output. 
Once trained, the Gaussian process emulators  provide a continuous picture of the parameter space.
and act as a fast surrogate to the full model:
they predicts model output at arbitrary points in parameter space with negligible computational cost.
This effectively removes most practical barriers and enables parameter calibration through standard techniques such as MCMC. Emulators have been successfully used to study a wide range of physical systems, including galaxy formation \cite{Gomez:2012ak} and heavy-ion collisions \cite{Novak:2013bqa,Pratt:2015zsa,Sangaline:2015isa,Bernhard:2015hxa,Bernhard:2016tnd}.

The final step in the parameter estimation method is to use MCMC to calibrate the model parameters to optimally reproduce experimental observables, thereby extracting probability distributions for the true values of the parameters.
According to Bayes' theorem, the probability for the true parameters $\x\st$ is
\begin{equation}
  P(\x\st|X,Y,\y\ex) \propto P(X,Y,\y\ex|\x\st) P(\x\st).
  \label{eq:bayes}
\end{equation}
The left-hand side is the \emph{posterior}: the probability of $\x\st$ given the design $X$, computed observables $Y$, and experimental data $\y\ex$.
On the right-hand side, $P(\x\st)$ is the \emph{prior} probability -- encapsulating initial knowledge of $\x\st$ -- and $P(X,Y,\y\ex|\x\st)$ is the likelihood: the probability of observing $(X, Y, \y\ex)$ given a proposal $\x\st$:
\begin{equation}
P(X,Y,\y\ex|\x\st)
    \propto\exp\left[
      -\frac{1}{2} (\y(\x\st) - \y\ex)\tran \Sigma_y^{-1} (\y(\x\st) - \y\ex) \right] \,,
\end{equation}
where $\y\ex$ is the experimental data and $\Sigma$ is the covariance matrix, which is the total of experimental statistical and systematic uncertainty, model statistical uncertainty and GP predictive uncertainty. For simplicity, we place a uniform prior on the model parameters, i.e.\ the prior is constant within the design range and zero outside.

\begin{figure}[t]
\begin{center}
\includegraphics[width=0.79\linewidth]{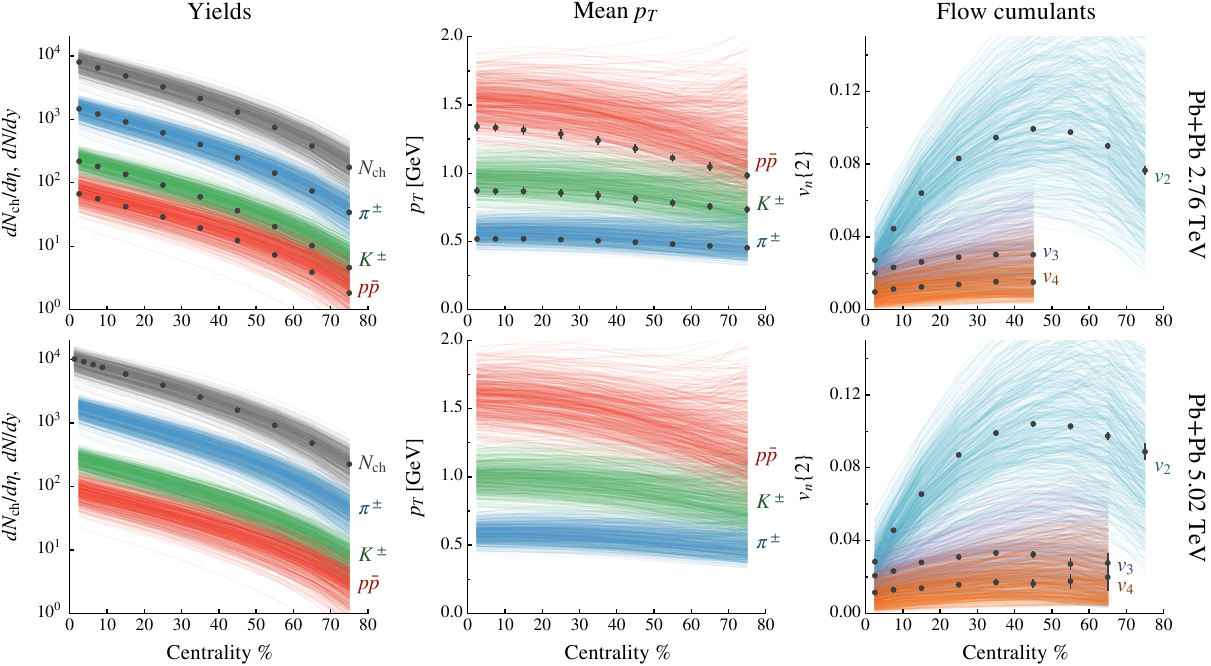}
\includegraphics[width=0.79\linewidth]{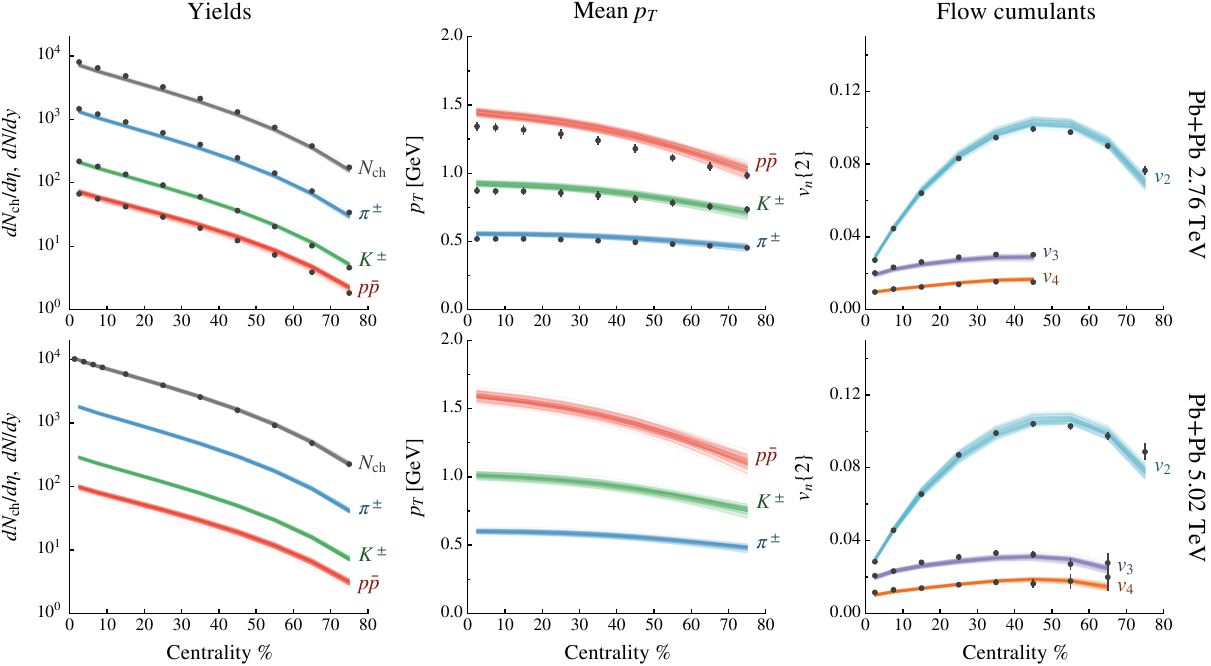}
\caption{\label{fig:prior-posterior}
Simulated observables compared to experimental data from the ALICE experiment \cite{Abelev:2013vea,ALICE:2011ab}.
  Top two rows: explicit model calculations for each of the 300 design points, for 2.76 and 5.02 TeV beam energy respectively;
  bottom two rows: emulator predictions of 100 random samples drawn from the posterior distribution.
  Left column: identified particle yields $dN/dy$,
  middle: mean transverse momenta $\avg{p_T}$,
  right: flow cumulants $v_n\{2\}$.
}
\end{center}
\end{figure}

Figure~\ref{fig:prior-posterior} compares simulated observables to experimental data and demonstrates the MCMC calibration. The top two rows have explicit model calculations at each of the 300 design points (chosen by a Latin hypercube algorithm) at which the actual physics model was evaluated. All model parameters vary across their full ranges, leading to the large spread in computed observables. The results of the MCMC calibration can be seen in the bottom two rows: these show emulator predictions of 100 random samples from the posterior distribution. Here, the model has been calibrated to the experiment, so its calculations are clustered tightly around the data -- although some uncertainty remains since the samples are drawn from a posterior distribution of finite width. Overall, the calibrated model provides an excellent simultaneous fit to all observables except for the mean $p_T$ of anti-protons (and to a lesser extent kaons), which is systematically overpredicted by 10\%.

\section{Determination of Initial Conditions and the Temperature-Dependence of the Shear Viscosity}

\begin{figure}[tb]
\begin{center}
\includegraphics[width=0.89\linewidth]{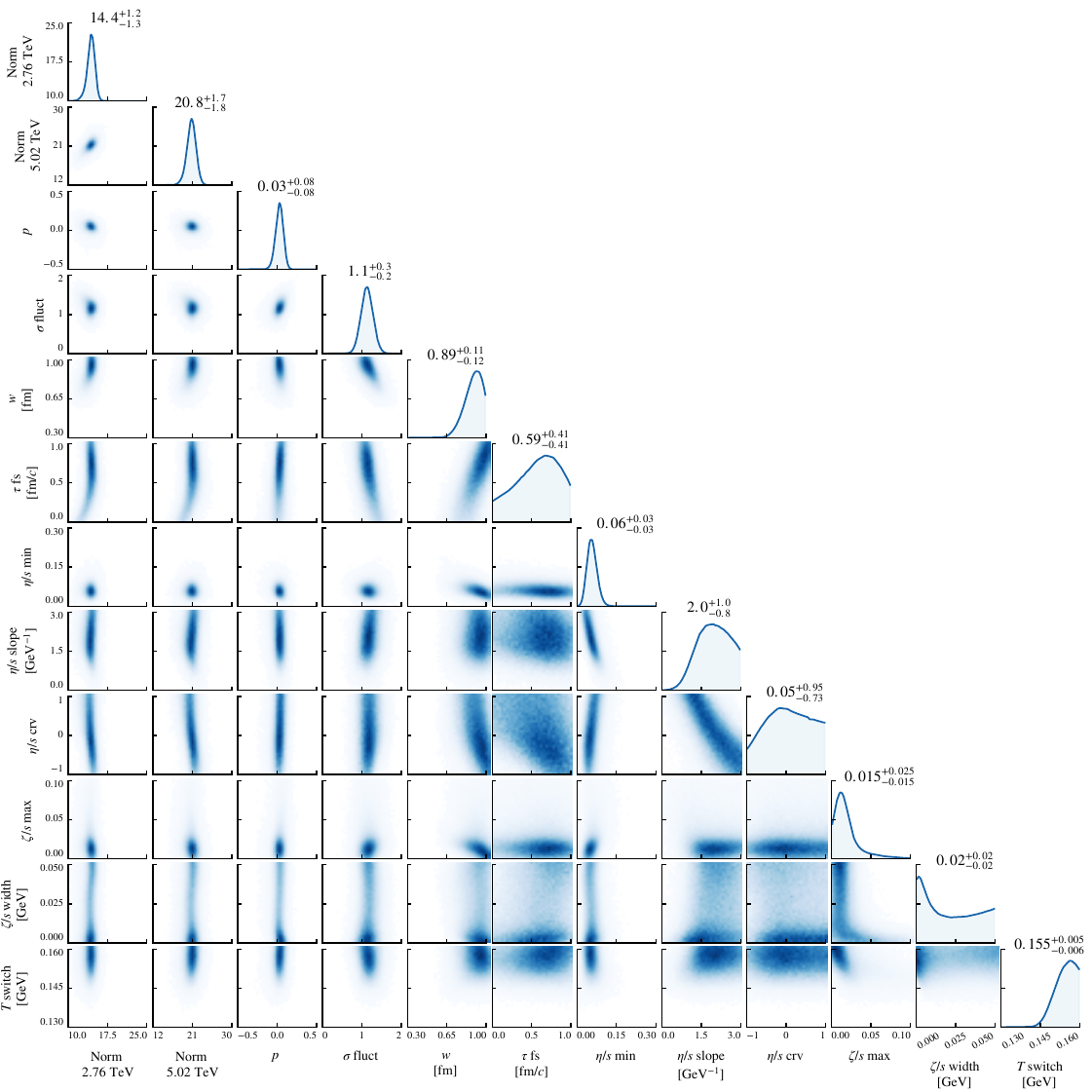}
\caption{\label{fig:posterior}
Posterior distributions for the model parameters.
  The diagonal has marginal distributions for each parameter, while the off-diagonal contains joint distributions showing correlations among pairs of parameters.
}
\end{center}
\end{figure}

The primary result of the analysis is the posterior distribution for the model parameters, Fig.~\ref{fig:posterior}. Here, the diagonal elements are marginal distributions for each model parameter (all other parameters integrated out), while the off-diagonals are joint distributions showing correlations among pairs of parameters. Operationally, these are all histograms of MCMC samples and the diagonal represents the individual probability distributions for all parameters of the model. One should note that the posterior distribution provides far more information than a simple least square fit for the extraction of optimum parameter values. Full probability distributions for all parameters as well as their pairwise correlations are given, enabling a rigorous assessment on how meaningful a particular parameter is for the physics model and how well the data actually constrains the values of the parameters.

\begin{figure}[htb]
\begin{center}
\includegraphics[width=0.75\linewidth]{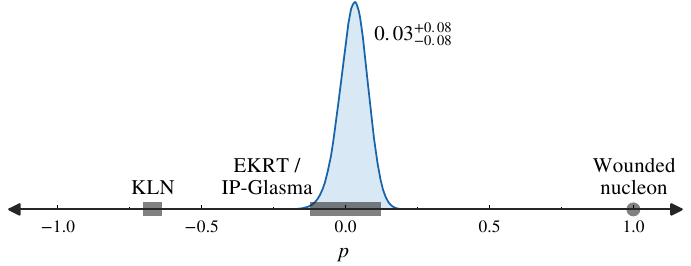}
\caption{\label{fig:posterior_p}
Posterior distribution of the \protect\trento\ entropy deposition parameter $p$.
  Approximate $p$-values are annotated for the KLN ($p~\approx~0.67~\pm~0.01$), EKRT ($p~\approx~0.0~\pm~0.1$), and wounded nucleon ($p = 1$) models.
}
\end{center}
\end{figure}

The first result we wish to highlight is the \trento\ entropy deposition parameter $p$, which  has a remarkably narrow distribution peaked at essentially zero with approximate 90\% uncertainty of $\pm0.1$. This implies that initial state entropy deposition is roughly proportional to the geometric mean of participant nuclear thickness functions, $s \sim \sqrt{\T_A\T_B}$.
This confirms our previous analysis of the \trento\ model which demonstrated that $p \approx 0$ simultaneously produces the correct ratio between initial state ellipticity and triangularity and fits multiplicity distributions for a variety of collision systems \cite{Moreland:2014oya}.
We observe little correlation between $p$ and any other parameters, suggesting that its optimal value is mostly factorized from the rest of the model.
Further, recall that the $p$ parameter smoothly interpolates among different classes of initial condition models;
Fig.~\ref{fig:posterior_p} shows an expanded view of the posterior distribution along with the approximate $p$-values for the other models: The EKRT and IP-Glasma models \cite{Niemi:2015qia,Schenke:2012wb} lie squarely in the peak -- this helps explain their success -- while the KLN and wounded nucleon models are considerably outside.

The shear viscosity parameters $(\eta/s)_\text{min,slope,curvature}$ set the temperature dependence of $\eta/s$ according to the ansatz
\begin{equation}
  (\eta/s)(T) = (\eta/s)_\text{min} + (\eta/s)_\text{slope} \, (T - T_c) \, (T - T_c)^{(\eta/s)_\text{curvature}}
  \label{eq:etas2}
\end{equation}
for $T > T_c$.
The full parametrization also includes a constant $(\eta/s)_\text{hrg}$ for $T < T_c$; this parameter was included in the calibration but yielded an essentially flat posterior distribution, implying that it has little to no effect.
This is not surprising, since hadronic viscosity is largely handled by UrQMD, not the hydrodynamic model.
Therefore, we omit $(\eta/s)_\text{hrg}$ from the posterior distribution visualizations.

Examining the marginal distributions for $\eta/s$ min, slope and curvature, we observe the distribution for $(\eta/s)_\text{min}$ (the value of $\eta/s$ at the transition temperature $T_c$ = 0.154 GeV) has a narrow peak at 0.06, below the KSS bound $1/4\pi \approx 0.08$ but consistent with it within 90\% uncertainty. On the other hand, zero $\eta/s$ is excluded at the 90\% level. The slope parameter has a broad peak, although zero slope is excluded, thus confirming a temperature-dependent (non-constant) $\eta/s$ at the 90\% level. The curvature parameter is not constrained, but exhibits a strong correlation with the slope. 

\begin{figure}[htb]
\begin{center}
\includegraphics[width=0.75\linewidth]{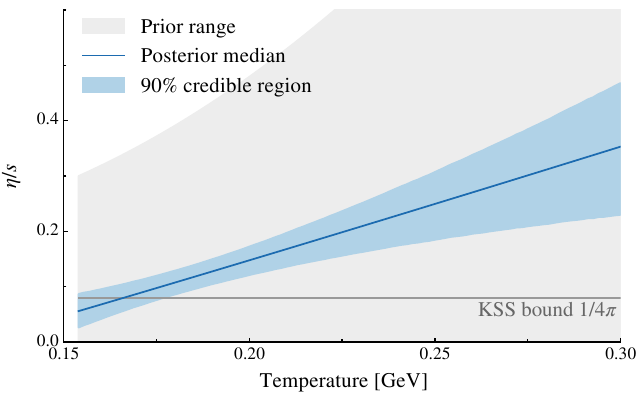}
\caption{\label{fig:etas_estimate}
Estimated temperature dependence of the shear viscosity $(\eta/s)(T)$ for $T > T_c = 0.154$ GeV.
  The gray shaded region indicates the prior range for the $(\eta/s)(T)$ parametrization,
  the blue line is the median from the posterior distribution,
  and the blue band is a 90\% credible region.
  The horizontal gray line indicates the KSS bound $\eta/s \geq 1/4\pi$ \cite{Danielewicz:1984ww, Policastro:2001yc, Kovtun:2004de}.
}
\end{center}
\end{figure}

Figure~\ref{fig:etas_estimate}  visualizes the estimated temperature dependence of $\eta/s$, including uncertainty, by inserting the posterior samples for the $\eta/s$ parameters back into Eq.~\ref{eq:etas2}. 
The figure reveals that the uncertainty is smallest at temperatures less than $T \approx {}$225 MeV.
We hypothesize that this is the most important temperature range for the present observables at LHC energies -- perhaps it is where the system spends most of its time and hence where most anisotropic flow develops, for instance -- and thus the data provide a ``handle'' for $\eta/s$ around 200~MeV.
Data at other beam energies and other, more sensitive observables could provide additional handles at different temperatures, enabling a more precise estimate of the temperature dependence of $\eta/s$ and possibly constrain the curvature in addition to the minimum and slope.

\section{Other Applications and Outlook}

Bayesian model-to-data comparisons have been well established and powerful tools in many different areas of science, e.g. in cosmology, particle physics and climate sciences. In relativistic heavy-ion physics they are fairly novel, having been pioneered by the MADAI collaboration \cite{Petersen:2010zt,Novak:2013bqa,Sangaline:2015isa,Bernhard:2015hxa}. One application of particular note is the extraction of the QCD equation of state probed in relativistic heavy-ion collisions using these techniques \cite{Pratt:2015zsa}. While the outcome of the analysis -- being compatible with Lattice field theory predictions -- may not come as a big surprise, it is nevertheless non-trivial, since it confirms the applicability of a Lattice equation of state calculated in the full equilibrium and infinite time limit for the short-lived mesoscopic system that constitutes a relativistic heavy-ion collision. At Quark Matter 2017, roughly half a dozen different presentations utilized Bayesian model-to-data comparisons for the extraction of a variety of different QGP properties -- ranging from initial conditions in pA systems \cite{Ke:QM17}, the sub-nucleonic structure of the initial state in AA collisions \cite{Moreland:QM17}, the temperature-dependence of the QGP shear- and bulk-viscosity \cite{Denicol:QM17,Bernhard:QM17} to the extraction of the heavy-quark transport coefficient \cite{Xu:QM17}.

Looking into the future, an analysis framework that can handle arbitrary numbers of inputs and outputs, systematically calculates quantitative constraints on all inputs simultaneously, and quickly evaluates the efficacy of physical models, will be of significant use for the field and enable multiple different quantitative studies and extractions of QGP properties.

The Duke QCD group acknowledges support by grants no. NSF-ACI-1550225 (NSF) and DE-FG02-05ER41367 (DOE). CPU time was provided by the Open Science Grid, supported by DOE and NSF, as well as the DOE funded National Energy Research Scientific Computing Center (NERSC).



%% References


%\bibliographystyle{h-physrev5}
%\bibliographystyle{elsarticle-num}
%\bibliography{/Users/bass/Github/Bibliography/Duke_QCD_refs}

\end{document}